\documentclass[12pt]{article}

\begin{document}

\begin{center}
    
{\bf {\Large Lorentz-Violating Supergravity}}

\bigskip\bigskip







Roland E. Allen \\ 
\smallskip
Center for Theoretical Physics \\
\smallskip
Texas A\&M University, College Station TX 77843, USA

\end{center}

\bigskip

\begin{abstract}

The standard forms of supersymmetry and supergravity are inextricably
wedded to Lorentz invariance. Here a Lorentz-violating form of 
supergravity is proposed. The superpartners have exotic properties 
that are not possible in a theory with exact Lorentz symmetry and 
microcausality. For example, the bosonic sfermions have spin 1/2 and the 
fermionic gauginos have spin 1. The theory is based on a phenomenological 
action that is shown to follow from a simple microscopic and statistical 
picture.
\end{abstract}

\section{Introduction}

There is no spin-statistics connection in nonrelativistic
quantum mechanics~\cite{allen1}, for a simple reason:
Although causality is a meaningful concept in
a nonrelativistic picture, microcausality is not. I.e., there is no
requirement that commutators vanish outside a light cone. For the same
reason, a fundamental theory which violates Lorentz invariance does not
necessarily lead to the usual spin-statistics connection~\cite{allen2}. Such
a theory may contain, e.g., bosons with spin 1/2 and fermions with spin
1, and these particles will have very unusual properties.
Here we will discuss a Lorentz-violating theory which contains spin 1/2
sfermions (bosonic superpartners of the Standard Model fermions), and spin
1 gauginos (fermionic superpartners of the Standard Model gauge bosons).

It is worthwhile first to clarify the vocabulary, in which the definition of
terms like ``supersymmetry'' is broadened in the interest of clarity~\cite
{encyclopedia}: An action or theory is defined to be supersymmetric if it is
invariant under a transformation which converts fermions to bosons and
vice-versa. Also, ``super'' is used as a generic prefix for any objects that
involve both commuting and anticommuting variables~\cite{allen3}. This usage
parallels that for ``complex'', which signifies that a quantity contains both
real and imaginary parts. The most standard forms of supersymmetry are
inextricably wedded to Lorentz invariance~\cite{kane}, whereas the broader
usage permits a Lorentz-violating theory to be supersymmetric. The fermionic
superpartners of Standard Model bosons still have the suffix ``ino'' (as in
gaugino) and the bosonic superpartners of fermions still have the prefix
``s'' (as in squark), although the ``s'' now stands for ``superpartner''
rather than ``scalar''.

Just as a theory can have Lorentz symmetry without supersymmetry, it can
have supersymmetry without Lorentz symmetry. Of course, a true theory of
nature must exhibit near-perfect Lorentz symmetry in those regimes where
this symmetry has been tested by experiment and observation.

There are several motivations for Lorentz-violating theories, ranging from
proposed \textit{ad hoc} solutions of problems in astrophysics~\cite{coleman}
to the anticipation that some mechanism may lead to a breaking of Lorentz
invariance that is experimentally 
detectable~\cite{colladay,kostelecky}. The present theory has a different
motivation: Namely, the goal is to understand \textit{why}
Lorentz invariance is a fundamental symmetry, how it emerges from a still 
more fundamental picture, and under what circumstances it may be violated. In
the second part of this paper (Sections 3 and 4), we will see that a simple
microscopic and statistical picture leads to the following phenomenological
and supersymmetric action:
\begin{equation}
S=\int d^{D}x\left[
\frac{1}{2m}\partial ^{M}\Psi ^{\dagger }\partial _{M}\Psi -\mu \Psi
^{\dagger }\Psi +\frac{1}{2}b\left( \Psi ^{\dagger }\Psi \right) ^{2}\right]
\end{equation}
\begin{equation}
\Psi =\left(
\begin{array}{c}
\Psi _{b} \\
\Psi _{f}
\end{array}
\right) \qquad
\end{equation}
where the notation is explained in Refs. 2 and 4.
In the first part of the paper (Section 2), we will start
with this action and demonstrate that it leads to a Lorentz-violating
form of supergravity. We will, however, emphasize
only the action for Standard Model fermions and their superpartners.
This action involves the coupling of the fermions and sfermions to the gauge
fields and gravity, and to gauginos and gravitinos, but the complete action
for the Standard Model bosons and their superpartners will be deferred
for future discussion elsewhere.

\section{Lorentz-Violating Supersymmetry and Supergravity}

After a transformation to Lorentzian spacetime, the Euclidean action of
(1) becomes
\begin{equation}
S_{L}=-\int d^{D}x\left[ -\frac{1}{2m}\Psi^{\dagger }\partial^{M}
\partial_{M}\Psi -\mu \Psi^{\dagger }\Psi +\frac{1}{2}b\left( \Psi
^{\dagger }\Psi \right) ^{2}\right] \, .
\end{equation}
Here, as in Refs. 2 and 4, the following convention is used for 
arbitrary D-dimensional vectors $a_{M}$: in the Euclidean formulation, 
$a^{M}=\delta^{MN}a_{N}$, but in the Lorentzian formulation, 
$a^{M}=\eta^{MN}a_{N}$, where $\eta^{MN}=diag(-1,1,...,1)$ is the 
Minkowski metric tensor in $D$ dimensions ($M=0,1,2,...,D-1$). (In 
both the Euclidean and Lorentzian formulations, $x^{0}$ is the 
coordinate corresponding to the physical time evolution of field densities 
like (51), and only the mathematical description is different.) The 
metric tensor associated with gravity and local Lorentz invariance, 
$g_{MN}$ or $g^{MN}$, is always shown explicitly in expressions like 
$g^{MN}a_{N}$.

As in Refs. 2 and 4, it is assumed that the physical vacuum contains a
GUT-scale condensate which forms in the very early universe, and which
exhibits rotations in both external spacetime $x^{\mu }$ ($\mu =0,1,2,3$)
and an internal space $x^{m}$ ($m=4,5,...,D-1$). These rotations are
described by unitary matrices $U_{ext}$ and $U_{int}$, so that the order
parameter
\begin{equation}
\Psi _{cond}=\left\langle \Psi _{b}\right\rangle _{vac}
\end{equation}
has the form
\begin{eqnarray}
\Psi _{cond} &=&\Psi _{ext}\left( x^{\mu }\right) \,\Psi _{int}\left(
x^{m},x^{\mu }\right) \\
\Psi _{ext}\left( x^{\mu }\right) &=&U_{ext}\left( x^{\mu }\right)
\,n_{ext}^{1/2}\left( x^{\mu }\right) \eta _{ext} \\
\Psi _{int} &=&U_{int} \left(x^{m},x^{\mu }\right)\,
n_{int}^{1/2}\left(x^{m},x^{\mu }\right) \eta _{int}
\end{eqnarray}
where $\eta _{ext}$ and $\eta _{int}$ are constant vectors.
External and internal ``superfluid velocities'' are defined by
\begin{eqnarray}
mv^{\mu } &=&-iU_{ext}^{-1}\partial ^{\mu }U_{ext}-iU_{int}^{-1}\partial
^{\mu }U_{int} \\
mv_{m} &=&-iU_{int}^{-1}\partial _{m}U_{int} \, .
\end{eqnarray}
Let us write
\begin{equation}
v^{M} = v_{\alpha }^{M}\sigma ^{\alpha }+v_{c}^{M}\sigma ^{c} \quad , \quad
v_{M} = v_{M\alpha }\sigma ^{\alpha }+v_{Mc}\sigma ^{c}
\end{equation}
where $\alpha = 0,1,2,3$ and $c>3$. Also, $\sigma ^{k}$ ($k=1,2,3$)
is a Pauli matrix, $\sigma ^{0}$ is the
$2\times 2$ identity matrix, and the $\sigma ^{c}$ are generators associated
with an initial internal symmetry group $G_{int}$. As in Refs. 2 and 4,
$v_{\alpha }^{\mu }$ is interpreted as the gravitational vierbein 
$e_{\alpha }^{\mu }$:
\begin{eqnarray}
e_{\alpha }^{\mu } &=&v_{\alpha }^{\mu } \\
g^{\mu \nu } &=&\eta ^{\alpha \beta }e_{\alpha }^{\mu }e_{\beta }^{\nu } \, .
\end{eqnarray}
Also, $v_{\mu c}$ is interpreted as giving the gauge potentials
$A_{\mu}^{i} $ through
\begin{equation}
e_{\mu c} = A_{\mu }^{i}K_{i}^{n}v_{nc}
\end{equation}
with
\begin{equation}
e_{\mu c} = -v_{\mu  c}
\end{equation}
where $K_{i}^{n}$ corresponds to the Killing vectors of the internal space,
just as in classic Kaluza-Klein theories.

In the present paper we wish to generalize the above treatment by
allowing for more general rotations which mix bosonic 
and fermionic degrees of freedom. First consider the global supersymmetry
transformation
\begin{equation}
\Psi \rightarrow \Psi ^{\prime }=\mathcal{U\,}\Psi
\end{equation}
or, with bosonic and fermionic fields shown separately,
\begin{equation}
\left(
\begin{array}{c}
\Psi _{b} \\
\Psi _{f}
\end{array}
\right) \rightarrow \left(
\begin{array}{c}
\Psi _{b}^{\prime } \\
\Psi _{f}^{\prime }
\end{array}
\right) =\left(
\begin{array}{ll}
\mathcal{U}_{bb} & \mathcal{U}_{bf} \\
\mathcal{U}_{fb} & \mathcal{U}_{ff}
\end{array}
\right) \left(
\begin{array}{c}
\Psi _{b} \\
\Psi _{f}
\end{array}
\right) .
\end{equation}
If
\begin{equation}
\mathcal{U\,}^{\dagger }\mathcal{U}=1
\end{equation}
(3) is invariant under this transformation, so the theory is
supersymmetric according to the definition given above. The elements
of $\mathcal{U}_{bb}$ and $\mathcal{U}_{ff}$ are ordinary commuting
variables, like the components of $\Psi _{b}$. The elements of
$\mathcal{U}_{bf}$ and $\mathcal{U}_{fb}$ are anticommuting Grassmann
variables, like the components of $\Psi _{f}$.

Now let us replace the picture of a rotating GUT-scale condensate by a more
general picture in which all the fields of the vacuum contain a rotation
described by a supermatrix $\mathcal{U}$ which varies as a function of 
the spacetime coordinates. With a possible redefinition of the fermion 
fields, we can choose $\mathcal{U}_{ff}=\mathcal{U}_{bb}$ and write
\begin{equation}
\Psi ^{vac} = \mathcal{U} \, n_{vac}^{1/2} \, \Psi ^{0}
\end{equation}
where $\Psi ^{0}$ is constant and
\begin{equation}
\Psi ^{vac} =\left\langle \Psi \right\rangle _{vac}
= \left(
\begin{array}{c}
\left\langle \Psi _{b}\right\rangle _{vac}  \\
\left\langle \Psi _{f}\right\rangle _{vac}
\end{array}
\right)
\end{equation}
\begin{equation}
\mathcal{U=}\left(
\begin{array}{ll}
\mathcal{U}_{bb} & \mathcal{U}_{bf} \\
\mathcal{U}_{fb} & \mathcal{U}_{bb}
\end{array}
\right)
\end{equation}
\begin{equation}
\Psi ^{0 \dagger } \Psi ^{0} = 1 \, .
\end{equation}

The generalizations of (8) - (10) and (13) - (14) are
\begin{eqnarray}
mV^{\mu } &=&-i\mathcal{U}^{-1}\partial ^{\mu }\mathcal{U} \\
\partial _{m}\mathcal{U} &=& 
\partial _{m} {U} = i \, U \, m \, v_{m} \\
V^{M} &=& V_{\alpha }^{M}\sigma ^{\alpha }+V_{c}^{M}\sigma ^{c} 
\quad , \quad
V_{M} = V_{M \alpha }\sigma ^{\alpha }+V_{Mc}\sigma ^{c} \\
E_{\mu c} &=& \mathcal{A}_{\mu }^{i}K_{i}^{n}v_{nc} \\
E_{\mu c} &=& -V_{\mu  c}
\end{eqnarray}
where the last two expressions in (23) implicitly multiply a $2 \times 2$ 
identity matrix, and it is assumed that the internal coordinate space 
contains no supersymmetric rotations.

The fact that $\mathcal{U}$ is unitary implies that $\partial _{M}\mathcal{U}
^{\dagger }\mathcal{U}=-\mathcal{U}^{\dagger }\partial _{M}\mathcal{U}$ with
$\mathcal{U}^{\dagger }=\mathcal{U}^{-1}$, or
\begin{equation}
mV_{M}=i\partial _{M}\mathcal{U}^{\dagger }\mathcal{U}
\end{equation}
so that
\begin{equation}
V_{M}^{\dagger }=V_{M} \quad , V^{ M \dagger }=V^{M} \quad \,.
\end{equation}
We can then write, e.g.,
\begin{equation}
V^{M}=\left(
\begin{array}{ll}
V_{bb}^{M} & V_{bf}^{M} \\
V_{bf}^{M \dagger } & V_{bb}^{M}
\end{array}
\right) \, .
\end{equation}

\bigskip At this point, the logic of Ref. 2 can be repeated with
\begin{equation}
v_{M} \rightarrow V_{M} \quad , \quad v^{M} \rightarrow V^{M}
\end{equation}
\begin{eqnarray}
e_{\alpha }^{\mu } &\rightarrow &E\,_{\alpha }^{\mu }=\left(
\begin{array}{ll}
e_{\alpha }^{\mu } & f_{\alpha }^{\mu } \\
f_{\alpha }^{\mu \dagger } & e_{\alpha }^{\mu }
\end{array}
\right) \\
A_{\mu }^{i} &\rightarrow &\mathcal{A}_{\mu }^{i}=\left(
\begin{array}{ll}
A_{\mu }^{i} & B_{\mu }^{i} \\
B_{\mu }^{i \dagger} & A_{\mu }^{i}
\end{array}
\right) \, .
\end{eqnarray}

In particular, (3.38) - (3.41) of Ref. 2 become
\begin{equation}
\Psi \left( x^{\mu },x^{m}\right) =\mathcal{U}\left( x^{\mu },x^{m}\right) 
\Psi^{r}\left( x^{\mu }\right) \psi _{r}^{int}\left( x^{m}\right) ,
\end{equation}
\begin{equation}
\partial _{\mu }\Psi = \mathcal{U}\left( x^{\mu },x^{m}\right) 
\left( \partial _{\mu }+imV_{\mu \alpha}\sigma ^{\alpha }
+imV_{\mu c}\sigma ^{c}\right) \Psi^{r}\psi _{r}^{int}
\end{equation}
\begin{eqnarray}
\int d^{d}x\,\Psi ^{\dagger }\,\partial ^{\mu }\partial _{\mu }\,\Psi
&=&\int d^{d}x\,\psi _{r}^{int\dagger }\Psi^{r\dagger
}\eta ^{\mu \nu }\left( \partial _{\mu }+imV_{\mu \alpha }\sigma ^{\alpha
}+imV_{\mu c}\sigma ^{c}\right)  \nonumber \\
&{}& \hspace{1.3cm} \times \left( \partial _{\nu }+imV_{\nu \beta
}\sigma ^{\beta }+imV_{\nu d}\sigma ^{d}\right) \Psi^{s}\psi _{s}^{int}
 \nonumber \\
&=&\Psi ^{r\dagger }\,\eta ^{\mu \nu }\langle r|\left(
\partial _{\mu }+imV_{\mu \alpha }\sigma ^{\alpha }+imV_{\mu c}\sigma
^{c}\right) \nonumber \\
&{}&  \hspace{1.3cm} \times \sum_{t}|t\rangle \langle t|\,\left( \partial _{\nu }
+imV_{\nu\beta }\sigma ^{\beta }+imV_{\nu d}\sigma ^{d}\right) |s\rangle \,
\Psi^{s}  \nonumber \\
&=&\Psi ^{r\dagger }\,\eta ^{\mu \nu }\left[ \delta
_{rt}\left( \partial _{\mu }+imV_{\mu \alpha }\sigma ^{\alpha }\right)
-i\mathcal{A}_{\mu }^{i}t_{i}^{rt}\right] \nonumber \\
&{}&  \hspace{1.3cm} \times \left[ \delta _{ts}\left( \partial
_{\nu }+imV_{\nu \beta }\sigma ^{\beta }\right) -i\mathcal{A}_{\nu
}^{j}t_{j}^{ts}\right] \,\Psi^{s}  \nonumber \\
&=&\Psi _{ext}^{\dagger }\,\eta ^{\mu \nu }\left[ \left(
\partial _{\mu }-i\mathcal{A}_{\mu }^{i}t_{i}\right) +imV_{\mu \alpha
}\sigma ^{\alpha }\right] \nonumber \\
&{}&  \hspace{1.5cm} \times \left[ \left( \partial _{\nu }-i\mathcal{A}_{\nu
}^{j}t_{j}\right) +imV_{\nu \beta }\sigma ^{\beta }\right] \,\Psi _{ext}
\end{eqnarray}
\begin{eqnarray}
S_{L}&=&\int d^{4}x\,\Psi _{ext}^{\dagger } \times \nonumber \\
&{}& \hspace{-1cm} \left( \frac{1}{2m}%
\mathcal{D}^{\mu }\mathcal{D}_{\mu }+\frac{1}{2}iV_{\alpha }^{\mu }\sigma
^{\alpha }\mathcal{D}_{\mu }+\frac{1}{2}\mathcal{D}_{\mu }iV_{\alpha }^{\mu
}\sigma ^{\alpha }-\frac{1}{2}mV^{\mu }_{\alpha }V_{\mu \alpha}+\mu
_{ext}\right) \Psi _{ext}
\end{eqnarray}
where
\begin{equation}
\mathcal{D}_{\mu }=\partial _{\mu }-i\mathcal{A}_{\mu }^{i}t_{i} \, .
\end{equation}

Also, the generalization of (2.22) of Ref. 2 is
\begin{equation}
\Psi ^{0\dagger }n_{vac}^{1/2}\left[ \left( \frac{1}{2}
mV^{\mu }V_{\mu }-\frac{1}{2m}\partial ^{\mu }\partial _{\mu }-\mu
_{ext}\right) -i\left( \frac{1}{2}\partial ^{\mu }V_{\mu }+V^{\mu }\partial
_{\mu }\right) \right] n_{vac}^{1/2}\Psi ^{0}=0 \, .
\end{equation}
Adding this equation to its Hermitian conjugate gives a still more 
general Bernoulli equation
\begin{equation}
\frac{1}{2}m\Psi ^{0\dagger }\,V^{\mu }V_{\mu }\,
\Psi ^{0}+P_{ext}=\mu _{ext}
\end{equation}
where
\begin{equation}
P_{ext}=-\frac{1}{2m}n_{vac}^{-1/2}\partial ^{\mu }\partial _{\mu
}n_{vac}^{1/2}.
\end{equation}

\noindent As before, it is assumed that the basic texture of the 
vacuum field rotations is such that
\begin{equation}
V_{0}^{k}=V_{a}^{0}=0 \quad , \quad k,a=1,2,3,
\end{equation}
and that the nonzero gauge potentials are not coupled to $\Psi ^{0}$ at 
energies well below the GUT scale~\cite{allen3}, so that (39) reduces to
\begin{equation}
{\frac{1}{2}}mV_{\alpha }^{\mu }V_{\alpha \mu }+P_{ext}=\mu _{ext}.
\end{equation}
When $\partial _{\mu }n_{vac}^{1/2}$ and $\partial ^{\mu }V_{\mu }$ 
are neglected, (36) then simplifies to
\begin{equation}
S_{L}=\int d^{4}x\,{}\Psi _{ext}^{\dagger }\left( \frac{1}{2m}\mathcal{D}^{\mu }
\mathcal{D}_{\mu }+iE_{\alpha }^{\mu }\sigma ^{\alpha }\mathcal{D}_{\mu
}\right) \Psi _{ext} \, .
\end{equation}
Since $m$ is
comparable to the Planck mass, it is reasonable to assume that the first
term can be neglected, giving
\begin{equation}
S_{L}=\int d^{4}x \, \Psi _{ext}^{\dagger } \, iE_{\alpha }^{\mu }\sigma
^{\alpha }\mathcal{D}_{\mu } \, \Psi _{ext}
\end{equation}
or, with $e_{\alpha }^{\mu }$ again slowly varying,
\begin{equation}
S_{L}=\int d^{4}x \, e \, \overline{\Psi }_{ext}^{\dagger } \, iE_{\alpha} 
^{\mu }\sigma ^{\alpha }\mathcal{D}_{\mu } \, \overline{\Psi }_{ext}
\end{equation}
\begin{equation}
\overline{\Psi }_{ext}=e^{-1/2}\,\Psi _{ext}\quad ,\quad e=\det \left(
e_{\alpha \mu }\right)\, .
\end{equation}

According to (31) and (32), the bosonic fields play the same role
as before. Namely, $e^{\mu}_{\alpha}$ is the vierbein representing
the gravitational field, and $A_{\mu}^{i}$ is the potential
representing the gauge fields of a grand-unified theory -- an $SO(10)$
theory~\cite{barger,mohapatra,klapdor} if the dimension of the
internal space is $10$. The fermionic fields can be interpreted in an
equally simple way: namely, $f^{\mu}_{\alpha}$ corresponds to a spin
2 gravitino and $B_{\mu}^{i}$ to spin 1 gauginos. Again, we have
generalized the usual vocabulary, so that the superpartner of the
graviton is defined to be the gravitino, and the superpartners of
gauge bosons to be gauginos, even though these fermions have quite
unconventional properties. The unconventional spin is required because,
e.g., $B_{\mu}^{i}$ transforms as a vector.

Here we have emphasized the fermions of the Standard Model and their
bosonic superpartners. We have shown that all of these particles are
automatically coupled to gauge fields, the gravitational field,
gauginos, and gravitinos, if one assumes the action (1) and gives
$A_{\mu}^{i}$,  $e^{\mu}_{\alpha}$, $B_{\mu}^{i}$, and $f^{\mu}_{\alpha}$
the physical interpretation above. The Higgs sector has not been
considered here; as will be discussed elsewhere, the action
for Higgs fields is interpreted as a low-energy effective action. 
As will also be discussed elsewhere, the action for the gauge 
fields, gravitational field, gauginos, and gravitinos is now interpreted 
as arising from a ``vacuum diamagnetism'', rather than just from the 
topological defects responsible for curvature~\cite{allen3}. 
Finally, a discussion of the detailed 
phenomenology (including the existence of a stable LSP) 
will again be presented elsewhere.

\section{Statistical Origin of the Bosonic Action}

In this section we turn to a different issue: the origin of the
phenomenological action (1). We will show that this action follows
from a simple microscopic and statistical picture. Our starting point
is a single fundamental system
which consists of $N_{w}$ identical ``whits'', with $N_{w}$ variable.
(``Whit'', whose meanings include ``particle'' and ``least possible
amount'', is an appropriate name for the irreducible objects that are
postulated here.) Each whit can exist in any of $M_{w}$ states, with the
number of whits in the $i$th state represented by $n_{i}$. A
\textit{microstate}
of the fundamental system is specified by the number of whits
and the state of each whit. A \textit{macrostate} is specified by only the
occupancies $n_{i}$ of the states.

As discussed below, $D$ of the states are used to define $D$ coordinates
$x^{M}$ in Euclidean spacetime, $m_{w}$ of the states are used to define
observable fields $\phi _{k}$, and the remaining $\left(
M_{w}-m_{w}-D\right) $ states may be regarded as corresponding to fields
that are unobservable (at least at the energy scales considered here).

Let us begin by defining an initial set of coordinates $X^{M}$ in terms of
the occupancies $n_{M}$:
\begin{equation}
X^{M}=\pm n_{M}a_{0}
\end{equation}
where $M=0,1,...,D-1$. It is convenient to include a fundamental length
$a_{0}$ in this definition, so that we can later express the coordinates in
conventional units. As one might expect, $a_{0}$ will eventually turn out to
be comparable to the Planck length:
\begin{equation}
a_{0}\sim \ell _{P}=\left( 16\pi G\right) ^{1/2}
\end{equation}
since $a_{0}=m^{-1}\sim m_{P}^{-1}=\ell _{P}$ according to (78).

With the definition (47), positive and negative coordinates correspond to
the same occupancies. There are two relevant facts, however, which
make this definition physically acceptable: First,
two points whose coordinates differ by a minus sign are typically
separated by cosmologically large distances. Second, and more
importantly, the fields $\phi_{k}$ need not return to their original
values when they are evolved, according to their equation of motion,
from points with positive coordinates to points with negative
coordinates. I. e., the field configurations described by the two
sets of points can be regarded as distinct, and in this sense the 
points themselves are distinct. The different branches of the field 
configuration are analogous to the branches of a 
multivalued function like $z^{1/2}$, which are taken to correspond to  
distinct Riemann sheets.

At extremely small distances, spacetime is discrete in the present
theory, with a finite interval between two adjacent points $X^{M}$ and
$X^{M}+\delta X^{M}$: $\delta X^{M}=a_{0}$. 
As in Section 2, the $X^{M}$ are divided into $4$ external coordinates
$X^{\mu }$ and $(D-4)$ internal coordinates $X^{m}$. In the internal space it
is natural to have variations on a length scale that is comparable to $\ell
_{P}.$ In external spacetime, on the other hand, we wish to consider fields
which vary much more slowly, and it is convenient to average over a more
physically meaningful length scale. Let us therefore consider a
$D$-dimensional rectangular box centered on a point $\bar{X}$, with $X^{M}$
ranging from $\bar{X}^{M}-a^{M}/2$ to $\bar{X}^{M}+a^{M}/2$. For the $\left(
D-4\right) $ coordinates of internal space, $a^{m}$ is taken to be the
original fundamental length $a_{0}$. For the $4$ coordinates of external
spacetime, $a^{\mu }$ is taken to be an arbitrary length $a$, and we will
find that the final form of the action is independent of this parameter.

In this coarse-grained picture, the density of whits in the $i$th state is
\begin{equation}
\rho _{i}\left( \bar{X}\right) =N_{i}/\Delta V\quad ,\quad i=1,2,...,M_{w}
\end{equation}
where
\begin{equation}
N_{i}=\sum_{X}n_{i}\left( X\right) \quad ,\quad \Delta
V=\prod_{M}a^{M}=a^{4}a_{0}^{D-4}
\end{equation}
and the values of $X$ are those lying within the box centered on $\bar{X}$.
Let
\begin{equation}
\phi _{k}^{2}=\rho _{k}\quad ,\quad k=1,2,...,m_{w}.
\end{equation}
The initial bosonic fields $\phi _{k}$ are then real (but defined only up to
a phase factor $\pm 1$).

Let $\bar{S}\left( \bar{X}\right) $ be the entropy of the single box at
$\bar{X}$ for a given set of densities $\rho _{i}$, as defined by $\bar{S}
\left( \bar{X}\right) =\log \,W\left( \bar{X} \right) $ (in units with
$k_{B}=\hbar =c=1$). Here $W\left( \bar{X} \right) $ is the total number of
microstates in this box at fixed $\rho _{i}$ or $N_{i}$:
$W\left( \bar{X}\right) =\mathcal{N}\left( \bar{X}\right)!/\Pi _{i}\,
N_{i}\left( \bar{X}
\right) !$, with
\begin{equation}
\mathcal{N}\left( \bar{X}\right) =\sum_{i}N_{i}\left( \bar{X}\right) .
\end{equation}
The total number of available microstates for all points $\bar{X}$ is $W=\Pi
_{\bar{X}}\,W\left( \bar{X}\right) $, so the total entropy is
\begin{eqnarray}
\bar{S} &=&\sum_{\bar{X}}\,\bar{S}\left( \bar{X}\right) \\
\bar{S}\left( \bar{X}\right) &=&\log \Gamma \left( \mathcal{N}\left( \bar{X}
\right) +1\right) -\sum_{i}\log \Gamma \left( N_{i}\left( \bar{X}\right)
+1\right) \,.
\end{eqnarray}
It follows that
\begin{eqnarray}
\frac{\partial \bar{S}}{\partial N_{i}\left( \bar{X}\right) } &=&\psi
\,\left( \mathcal{N}\left( \bar{X}\right) +1\right) -\psi \left( N_{i}\left(
\bar{X}\right) +1\right) \\
\frac{\partial ^{2}\bar{S}}{\partial N_{i^{\prime }}\left( \bar{X} \right)
\partial N_{i}\left( \bar{X}\right) } &=&\psi \,^{\left( 1\right) }\left(
\mathcal{N}\left( \bar{X}\right) +1\right) -\psi ^{\left( 1\right) }\left(
N_{i}\left( \bar{X}\right) +1\right) \delta _{i^{\prime }i}
\end{eqnarray}
where $\psi \,\left( z\right) =d\log \Gamma \left( z\right) /dz$ and $\psi
^{\left( n\right) }\,\left( z\right) =d^{n+1}\log \Gamma \left( z\right)
/dz^{n+1}$ are the digamma and polygamma functions, with the asymptotic
expansions~\cite{abramowitz}
\begin{eqnarray}
\psi \,\left( z\right) &=&\log z-\frac{1}{2z}-\sum_{l=1}^{\infty }\frac{
B_{2l}}{2l\,z^{2l}}\quad \\
\psi ^{\left( n\right) }\,\left( z\right) &=&\left( -1\right) ^{n-1}\left[
\frac{\left( n-1\right) !}{z^{n}}+\frac{n!}{2z^{n+1}}+\sum_{l=1}^{\infty
}B_{2l}\frac{\left( 2l+n-1\right) !}{\left( 2l\right) !\,z^{n+2l}}\right]
\end{eqnarray}
as $z\rightarrow \infty .$ For $a\gg \ell _{P}$, we have $\mathcal{N}\left(
\bar{X}\right) >>>\bar{n}_{\mu }=\left( \bar{X}^{\mu }/a_{0}\right) ^{2}>>>1$
, so it is an extremely good approximation to write
\begin{eqnarray}
\frac{\partial \bar{S}}{\partial N_{k}\left( \bar{X}\right) } &=&\log
\mathcal{N}\left( \bar{X}\right) -\psi \left( N_{k}\left( \bar{X}\right)
+1\right) \\
\frac{\partial ^{2}\bar{S}}{\partial N_{k^{\prime }}\left( \bar{X} \right)
\partial N_{k}\left( \bar{X}\right) } &=&-\psi ^{\left( 1\right) }\left(
N_{k}\left( \bar{X}\right) +1\right) \delta _{k^{\prime }k}.
\end{eqnarray}

We could express $\,\bar{S}$ as a Taylor series expansion about the bare
vacuum with $\phi _{k}\left( \bar{X}\right) =0$ for all $k$ and $\bar{X}$:
\begin{eqnarray}
\bar{S} &=&S_{bare}+\sum_{\bar{X},k}\sum_{n}b_{n}\left( \bar {X}\right)
N_{k}\left( \bar{X}\right) ^{n} \\
b_{1}\left( \bar{X}\right) &=&\log \mathcal{N}_{bare}\left( \bar{X} \right)
-\psi \,\left( 1\right) \\
b_{n+1} &=&-\psi ^{\left( n\right) }\,\left( 1\right) /n!\quad ,\quad
n=1,2,...
\end{eqnarray}
with
\begin{eqnarray}
\psi \,\left( 1\right) &=&-\gamma \quad ,\quad \gamma =
\mbox{ Euler's
constant} \\
\psi ^{\left( n\right) }\,\left( 1\right) &=&\left( -1\right)
^{n+1}n!\,\zeta \left( n+1\right)
\end{eqnarray}
where $\mathcal{N}_{bare}\left( \bar{X}\right) $ is the value of $\mathcal{N}
\left(\bar{X}\right) $ when $N_{k}\left( \bar{X}\right) =0$ for all the
observable states $k$ and $\zeta \left( z\right) $ is the Riemann zeta
function. This is not physically appropriate, however, because bosonic
fields exhibit extremely large zero-point fluctuations in the physical
vacuum~\cite{sakurai}. (These are analogous to the zero-point oscillations
$\left\langle x^{2}\right\rangle $ of a harmonic oscillator, but with a very
large number of modes extending up to a Planck-scale cutoff.) In fact, it is
consistent with both standard physics and the treatment of this paper to
assume that
\begin{equation}
\left\langle \phi _{k}^{2}\right\rangle _{vac}=\left\langle \rho
_{k}\right\rangle _{vac}=\left\langle N_{k}\right\rangle _{vac}/\Delta V\sim
\ell _{P}^{-D}.
\end{equation}
Since there is no initial distinction between the states $\phi _{k}$, it is
reasonable to perform a Taylor series expansion about the same value $
N_{vac} $ for each $k$, where
\begin{equation}
N_{vac}\sim \ell _{P}^{-D}\Delta V\sim \left( a/\ell _{P}\right) ^{4}>>>1
\end{equation}
if, e.g., $a^{-1}\sim 10^{10}$ TeV (with $\ell _{P}^{-1}=m_{P}\sim 10^{15}$
TeV). It is then an extremely good approximation to use the asymptotic
formulas above and write
\begin{equation}
\bar{S} =S_{vac}+\sum_{\bar{X},k}a_{1}\Delta N_{k}\left( \bar{X}\right)
+\sum_{\bar{X},k}a_{2}\left[ \Delta N_{k}\left( \bar{X}\right) \right] ^{2}
\end{equation}
\begin{equation}
\Delta N_{k}\left( \bar{X}\right) = N_{k}\left( \bar{X}\right) -N_{vac}
\end{equation}
\begin{equation}
a_{1} = \log \mathcal{N}_{vac}-\log N_{vac} \quad , \quad a_{2} = -1 /
\left( 2N_{vac} \right)
\end{equation}
where $\mathcal{N}_{vac}\left( \bar{X}\right) $ is the value of $\mathcal{N}
\left( \bar{X}\right) $ when $N_{k}\left( \bar{X}\right) =N_{vac}$ for all
$k $, and the neglected terms are of order $\left[ \Delta N_{k}\left( \bar{X}
\right) /N_{vac}\right] ^{n}\Delta N_{k}\left( \bar{X} \right) $, $n\geq 2$.

It is not conventional or convenient to deal with $\Delta N_{k}$ and $\left(
\Delta N_{k}\right) ^{2}$, so let us instead write $\bar{S}$ in terms of the
fields $\phi _{k}$ and their derivatives $\partial \phi _{k}/\partial x^{M}$
via the following procedure: First, we can switch from the original points
$\bar{X}$, which are defined to be the centers of the boxes, to a new set of
points $\widetilde{X}$, which will be defined to be the corners of the
boxes. It is easy to see that
\begin{equation}
\bar{S}=S_{vac}+\sum_{\widetilde{X},k}a_{1}\left\langle \Delta N_{k}\left(
\bar{X}\right) \right\rangle +\sum_{\widetilde{X},k}a_{2}\left\langle \left[
\Delta N_{k}\left( \bar{X}\right) \right] ^{2}\right\rangle
\end{equation}
where $\left\langle \cdots \right\rangle $ in the present context indicates
an average over the $2^{D}$ boxes labeled by $\bar{X}$ which have the common
corner $\widetilde{X}$. Second, we can write $\Delta N_{k}=\Delta \rho
_{k}\Delta V=\left( \left\langle \Delta \rho _{k}\right\rangle +\delta \rho
_{k}\right) \Delta V$, with $\left\langle \delta \rho _{k}\right\rangle =0$:
\begin{eqnarray}
\hspace{-0.5cm} \bar{S} &=&S_{vac}+
\sum_{\widetilde{X},k}a_{1}\left\langle \left\langle
\Delta \rho _{k}\right\rangle +\delta \rho _{k}\right\rangle \Delta V
+\sum_{\widetilde{X},k}a_{2}\left\langle \left( \left\langle \Delta \rho
_{k}\right\rangle +\delta \rho _{k}\right) ^{2}\right\rangle \left( \Delta
V\right) ^{2} \\
\hspace{-0.5cm} &=&S_{vac}+\sum_{\widetilde{X},k}a_{1}\left\langle \Delta \rho
_{k}\right\rangle \Delta V+\sum_{\widetilde{X},k}a_{2}\left[ \left\langle
\Delta \rho _{k}\right\rangle ^{2}+\left\langle \left( \delta \rho
_{k}\right) ^{2}\right\rangle \right] \left( \Delta V\right) ^{2}.
\end{eqnarray}
Each of the $2^{D}$ points $\bar{X}$ surrounding $\widetilde{X}$ is
displaced by $\pm a/2$ along the $x^{\mu }$ axes and $\pm a_{0}/2$ along the
$x^{m}$ axes. The last term above can therefore be rewritten
\begin{eqnarray}
\left\langle \left( \delta \rho _{k}\right) ^{2}\right\rangle &=&\sum_{\mu
}\left( \frac{\partial \rho _{k}}{\partial X^{\mu }}\right) ^{2}\left( \frac{
a}{2}\right) ^{2}+\sum_{m}\left( \frac{\partial \rho _{k}}{\partial X^{m}}
\right) ^{2}\left( \frac{a_{0}}{2}\right) ^{2} \\
&=&\sum_{\mu }\rho _{k}\left( \frac{\partial \phi _{k}}{\partial X^{\mu }}
\right) ^{2}a^{2}+\sum_{m}\rho _{k}\left( \frac{\partial \phi _{k}}{\partial
X^{m}}\right) ^{2}a_{0}^{2}
\end{eqnarray}
where the neglected terms involve higher derivatives and higher powers of $a$
and $a_{0}$. Since $\rho _{k}=\rho _{vac}+\Delta \rho _{k}$, with $\Delta
\rho _{k}<<<$ $\rho _{vac}=N_{vac}/\Delta V$ for normal fields, it is an
extremely good approximation to replace $\rho _{k}$ by $\rho _{vac}$ in the
above expression, and to neglect the term involving $a_{2}\left( \Delta
V\right) ^{2}\left( \Delta \rho _{k}\right) ^{2}=-\left( \Delta N_{k}\right)
^{2}/2N_{vac}$, so that we have
\begin{equation}
\bar{S}=S_{vac}^{\prime }+\sum_{\widetilde{X},k}\Delta V\,\left\{ \mu \bar{
\phi}_{k}^{2}-\frac{1}{2m}\,\left[ \sum_{\mu }\left( \frac{\partial \bar{\phi
}_{k}}{\partial X^{\mu }}\right) ^{2}\left( \frac{a}{a_{0}}\right)
^{2}+\sum_{m}\left( \frac{\partial \bar{\phi}_{k}}{\partial X^{m}}\right)
^{2}\right] \right\}
\end{equation}
where
\begin{equation}
m=a_{0}^{-1}\quad ,\quad \mu =m\left( \log \mathcal{N}_{vac}-\log
N_{vac}\right) \quad ,\quad \bar{\phi}_{k}=\phi _{k}/m
\end{equation}
and $S_{vac}^{\prime }=S_{vac}-\sum_{\widetilde{X},k}N_{vac}\left( \log
\mathcal{N}_{vac}-\log N_{vac}\right) $. Recall that
\begin{equation}
m\sim m_{P}=\ell _{P}^{-1}.
\end{equation}

The philosophy behind the above treatment is simple: We essentially wish to
replace $\left\langle f^{2}\right\rangle $ by $\left( \partial f/\partial
x\right) ^{2}$, and this can be accomplished because
\begin{equation}
\left\langle f^{2}\right\rangle -\left\langle f\right\rangle
^{2}=\left\langle \left( \delta f\right) ^{2}\right\rangle \approx
\left\langle \left( \partial f/\partial x\right) ^{2}\left( \delta x\right)
^{2}\right\rangle =\left( \partial f/\partial x\right) ^{2}\left( a/2\right)
^{2}.
\end{equation}
The form of (76) also has a simple interpretation: The entropy $\bar{S}$
increases with the number of whits, but decreases when the whits are not
uniformly distributed.

In the continuum limit,
\begin{equation}
\sum_{\widetilde{X}}\Delta V=\sum_{\widetilde{X}}a^{4}a_{0}^{D-4}\rightarrow
\int d^{D}X=\int_{a}^{\infty }d^{4}X\,\int_{a_{0}}^{\infty }d^{D-4}X
\end{equation}
(76) becomes
\begin{eqnarray}
\bar{S} &=&S_{vac}^{\prime }+\int_{a}^{\infty
}d^{4}X\,\int_{a_{0}}^{\infty }d^{D-4}X\,\sum_{k}  \\
&{}& \hspace{2cm} \times \left\{ \mu \bar{\phi}
_{k}^{2}-\,\frac{1}{2m}\left[ \sum_{\mu }\left( \frac{\partial \bar{\phi}_{k}
}{\partial X^{\mu }}\right) ^{2}\left( \frac{a}{a_{0}}\right)
^{2}+\sum_{m}\left( \frac{\partial \bar{\phi}_{k}}{\partial X^{m}}\right)
^{2}\right] \right\}  \nonumber \\
&=&S_{vac}^{\prime }+\int_{a_{0}}^{\infty }d^{D}x\,\,\sum_{k}\left[ \mu \Phi
_{k}^{2}-\,\frac{1}{2m}\,\sum_{M}\left( \frac{\partial \Phi _{k}}{\partial
x^{M}}\right) ^{2}\right]
\end{eqnarray}
where
\begin{equation}
x^{m}=X^{m}\quad ,\quad x^{\mu }=\left( a_{0}/a\right) X^{\mu }\quad ,\quad
\Phi _{k}=\left( a_{0}/a\right) ^{2}\bar{\phi}_{k}.
\end{equation}
The lower limit on each integral is the cutoff imposed by the size of the
rectangular boxes used in the coarse-graining above: $a$ for $X^{\mu }$,
$a_{0}$ for $X^{m}$, and $a_{0}$ for any $x^{M}$. The
continuum limit is an extremely good approximation for slowly varying
fields in external spacetime, but only a moderately good approximation
within the internal space, where the order parameter varies on a
length scale comparable to $\ell _{P}$. This implies that terms involving
higher derivatives $\partial ^{n}\widetilde{\phi }_{k}/\partial \left(
x^{m}\right) ^{n}$ can be significant in the internal space.

Notice that the final form (82) is independent of the arbitrary length $a$
which was used for coarse-graining in external spacetime. The fields must be
rescaled as $a$ is varied, but this is already a familiar feature in
standard physics~\cite{peskin}.

A physical configuration of all the fields $\phi _{k}\left( x\right) $
corresponds to a specification of all the densities $\rho _{k}\left(
x\right) $. In the present picture, the probability of such a configuration
is proportional to $W=e^{\bar{S}}$. In the Euclidean path integral, the
probability is proportional to $e^{-S_{E}}$, where $S_{E}$ is the Euclidean
action. We conclude that
\begin{equation}
S_{E}=-\bar{S}+\mbox{constant}.
\end{equation}
Choosing the constant to be zero, and employing the Einstein summation
convention for all repeated indices, we obtain
\begin{equation}
S_{E}=-S_{vac}^{\prime }+\int d^{D}x\,\left( \frac{1}{2m}\frac{\partial \Phi
_{k}}{\partial x_{M}}\frac{\partial \Phi _{k}}{\partial x^{M}}-\mu \Phi
_{k}\Phi _{k}\right) .
\end{equation}

The above result neglects interactions among the observable and
unobservable fields, which will arise from the 
higher-order terms neglected above. 
Since a detailed treatment of these interactions
would be quite complicated, we resort at this point to a phenomenological
description: We assume that probability can flow
out of and into each field, and that this effect can be modeled
by a random optical potential $i\widetilde{V}$ which has a Gaussian
distribution, with
\begin{equation}
\left\langle \widetilde{V}\,\right\rangle =0 \quad , \quad
\left\langle \widetilde{V}\left( x\right) \widetilde{V}\left( x^{\prime}
\right) \right\rangle =b\,\delta \left( x-x^{\prime }\right)
\end{equation}
where $b$ is a constant.

If we also assume that the number of observable real fields $\Phi
_{k}$ is even, we can group them in pairs to form complex fields $\Psi
_{b,k} $. Then we finally
have
$S_{E} = S_{0} +\bar{S}_{E}\left[ \Psi _{b},\Psi
_{b}^{\dagger }\right]$ with
\begin{equation}
\bar{S}_{E}\left[ \Psi _{b},\Psi _{b}^{\dagger }\right] = \int
d^{D}x\,\left( \frac{1}{2m}\partial ^{M}\Psi _{b}^{\dagger }\partial
_{M}\Psi _{b}-\mu \,\Psi _{b}^{\dagger }\Psi _{b}+i\widetilde{V}\,\Psi
_{b}^{\dagger }\Psi _{b}\right)
\end{equation}
where $\Psi _{b}$ is the vector with components $\Psi _{b,k}$.

\section{Supersymmetric Action}

If $F$ is a physical quantity which is determined by the observable fields,
its average value is given by
\begin{equation}
\left\langle F\right\rangle =\left\langle \frac{\int \mathcal{D}\,\Psi _{b}\,
\mathcal{D}\,\Psi _{b}^{\dagger }\,F\left[ \Psi _{b},\Psi _{b}^{\dagger
}\right] \,e^{-\bar{S}_{E}\left[ \Psi _{b},\Psi _{b}^{\dagger }\right] }}
{\int \mathcal{D}\,\Psi _{b}^{\prime }\,\mathcal{D}\,\Psi _{b}^{\prime
\dagger }\,e^{-\bar{S}_{E}\left[ \Psi _{b}^{\prime },\Psi _{b}^{\prime
\dagger }\right] }}\right\rangle
\end{equation}
where $\left\langle \cdot \cdot \cdot \right\rangle $ represents an average
over the perturbing potential $i \widetilde{V}\,$. The presence of the
denominator makes it difficult to perform this average, but there is a trick
for removing the bosonic degrees of freedom $\Psi _{b}^{\prime }$ in the
denominator and replacing them with fermionic degrees of freedom $\Psi _{f}$
in the numerator~\cite{parisi,efetov,huang,allen4}: Since
\begin{equation}
\int \mathcal{D}\,\Psi _{b}^{\prime }\,\mathcal{D}\,\Psi _{b}^{\prime
\dagger }\,e^{-\bar{S}_{E}\left[ \Psi _{b}^{\prime },\Psi _{b}^{\prime
\dagger }\right] }=\left( \det \,A\right) ^{-1}
\end{equation}
\begin{equation}
\int \mathcal{D}\,\Psi _{f}\,\mathcal{D}\,\Psi _{f}^{\dagger }\,e^{-\bar{S}
_{E}\left[ \Psi _{f},\Psi _{f}^{\dagger }\right] }=\det \,A
\end{equation}
where $A$ represents the operator of (87), it follows that
\begin{eqnarray}
\left\langle F\right\rangle &=&\left\langle \int \mathcal{D}\,\Psi _{b}\,
\mathcal{D}\,\Psi _{b}^{\dagger }\,\mathcal{D}\,\Psi _{f}\,\mathcal{D}\,\Psi
_{f}^{\dagger }\,F\,e^{-\bar{S}_{E}\left[ \Psi _{b},\Psi _{b}^{\dagger
}\right] }e^{-\bar{S}_{E}\left[ \Psi _{f},\Psi _{f}^{\dagger }\right]
}\right\rangle \\
&=&\left\langle \int \mathcal{D}\,\Psi \,\mathcal{D}\,\Psi ^{\dagger
}\,F\,e^{-\bar{S}_{E}\left[ \Psi ,\Psi ^{\dagger }\right] }\right\rangle
\end{eqnarray}
where $\Psi _{b}$ and $\Psi _{f}$ have been combined into $\Psi $,
\begin{equation}
\Psi =\left(
\begin{array}{c}
\Psi _{b} \\
\Psi _{f}
\end{array}
\right) ,
\end{equation}
and
\begin{equation}
\bar{S}_{E}\left[ \Psi ,\Psi ^{\dagger }\right] =\int d^{D}x\left[ \partial
^{M}\Psi ^{\dagger }\partial _{M}\Psi -\mu \Psi ^{\dagger }\Psi +i\widetilde{
V}\,\Psi ^{\dagger }\Psi \right] .
\end{equation}
(In (93), $\Psi _{f}$ consists of Grassmann variables
$\Psi _{f,k}$, just
as $\Psi _{b}$ consists of ordinary variables $\Psi _{b,k}$.) For a Gaussian
random variable $v$ whose mean is zero, the result
\begin{equation}
\left\langle e^{-iv}\right\rangle =e^{-\frac{1}{2}\left\langle
v^{2}\right\rangle }
\end{equation}
implies that
\begin{eqnarray}
\hspace{-1cm}\left\langle e^{-\int d^{D}x\, i\widetilde{V}\,\Psi ^{\dagger }\Psi
}\right\rangle &=&e^{-\frac{1}{2}\int d^{D}x\,\,\,d^{D}x\,^{\prime }\,\Psi
^{\dagger }\left( x\right) \Psi \left( x\right) \left\langle \widetilde{V}
\left( x\right) \widetilde{V}\left( x^{\prime }\right) \right\rangle \Psi
^{\dagger }\left( x^{\prime }\right) \Psi \left( x^{\prime }\right) } \\
&=&e^{-\frac{1}{2}b\int d^{D}x\,\,\,\left[ \Psi ^{\dagger }\left( x\right)
\Psi \left( x\right) \right] ^{2}}.
\end{eqnarray}
It follows that
\begin{equation}
\left\langle F\right\rangle =\int \mathcal{D}\,\Psi \,\mathcal{D}\,\Psi
^{\dagger }\,F\,e^{-S}
\end{equation}
with
\begin{equation}
S=\int d^{D}x\left[ \frac{1}{2m}\partial ^{M}\Psi ^{\dagger }\partial
_{M}\Psi -\mu \Psi ^{\dagger }\Psi +\frac{1}{2}b\left( \Psi ^{\dagger }\Psi
\right) ^{2}\right] .
\end{equation}
A special case is
\begin{equation}
Z=\int \mathcal{D}\,\Psi \,\mathcal{D}\,\Psi ^{\dagger }e^{-S}
\end{equation}
but according to (88)
\begin{equation}
Z=1.
\end{equation}
To make the expression for $\left\langle F\right\rangle $ independent of how
the measure is defined in the path integral, we can rewrite (98) as
\begin{equation}
\left\langle F\right\rangle =\frac{1}{Z}\int \mathcal{D}\,\Psi \,\mathcal{D}
\,\Psi ^{\dagger }\,F\,e^{-S}.
\end{equation}

Notice that the fermionic variables $\Psi _{f}$ represent true degrees of
freedom, and that they originate from the bosonic variables $\Psi
_{b}^{\prime }$. The coupling between the fields $\Psi _{b}$ and $\Psi _{f}$
(or $\Psi_{b}^{\prime }$) is due to the random perturbing potential
$i \widetilde{V}$.

\section{Conclusion}

The goal of this paper, and of the program represented by Refs. 2 and 4,
is ambitious: It is to start with a simple and convincing microscopic
picture, to show that this picture reproduces standard physics in the
appropriate regime (of known particles at energies far below the Planck
scale), and to propose extensions of standard physics that
are experimentally testable.

A truly fundamental theory should aspire to explaining the origins of

$\bullet$ Lorentz invariance

$\bullet$ bosonic fields

$\bullet$ fermionic fields

$\bullet$ supersymmetry

$\bullet$ gauge fields and their symmetry

$\bullet$ gravity

$\bullet$ quantum mechanics

$\bullet$ spacetime.

\bigskip

In the present paper, Lorentz invariance emerges for fermions
at energies that are
far below the Planck scale, in part because the second-order term in
(43) can be neglected at low energy. Bosonic fields, fermionic fields,
and supersymmetry emerge via the arguments of Sections 3 and 4,
immediately above. Gauge fields and
gravity, and well as gaugino and gravitino fields, emerge as
collective modes of the vacuum, described by the supermatrix $\mathcal{U}$
of (20). Quantum mechanics emerges through the chain of reasoning
which begins in Section 3: Starting with a statistical picture, 
one obtains an entropy $\bar{S}$ which is then
essentially interpreted as the negative of the Euclidean action,
according to (84). After a transformation to Lorentzian time, and a 
change from path-integral quantization to canonical quantization, one
obtains the usual formulation of quantum field theory. Finally,
spacetime emerges according to the definition (47) of initial spacetime
coordinates. Essentially, each coordinate $x^{M}$ corresponds to a different
state labeled by $M$, and the occupancy of this state measures the
position along the $x^{M}$ axis.

There are many obvious philosophical issues involved in, e.g., the
transition from Euclidean to Lorentzian time, and the identification
of occupancies with spacetime coordinates. It appears, nevertheless,
that these issues can be resolved and that the picture of the above
paragraph makes physical sense.

Most importantly, there are predictions that should be experimentally
testable within the next few years. In particular, the simplest form 
of the present theory 
predicts that each particle and its superpartner have the same spin,
so that bosonic sfermions have spin 1/2 and fermionic gauginos have
spin 1.


\end{document}